\documentstyle[prb,aps,tighten,preprint]{revtex}

\begin{document}
\title{Transition from insulating to a non-insulating temperature dependence of the
conductivity in granular metals}
\author{K.B. Efetov$^{1,2}$ and A. Tschersich$^{1}$}
\address{$^{1}$Fakult\"{a}t f\"{u}r Physik und Astronomie, Ruhr-Universit\"{a}t\\
Bochum, Universit\"{a}tsstr. 150, Bochum, Germany\\
$^{2}$L.D. Landau Institute for Theoretical Physics, Moscow, Russia}
\date{\today{}}
\maketitle

\begin{abstract}
We consider interaction effects in a granular normal metal at not very low
temperatures. Assuming that all weak localization effects are suppressed by
the temperature we replace the initial Hamiltonian by a proper functional of
phases and study the possibility for a phase transition depending on the
tunneling conductance $g$. It is demonstrated for any dimension that, while
at small $g$ the conductivity decays with temperature exponentially, its
temperature dependence is logarithmic at large $g.$ The formulae obtained
are compared with an existing experiment and a good agreement is found.
\end{abstract}


\draft

In an experiment \cite{Gerber97} on granular Al-Ge thick films several
interesting effects have been discovered. Destroying superconductivity by a
magnetic field up to $17T$ the authors could study, in particular,
properties of the normal state. Some features of the normal state related to
a negative magnetoresistance due to superconducting fluctuations have been
discussed recently \cite{bel} but an unusual observation remained completely
unexplained.

What we have in mind is a peculiar temperature dependence of the
conductivity found in some samples. Samples that had a high room temperature
resistivity showed an exponential decay of the conductivity as a function of
temperature. This behavior is typical for insulators and has been
interpreted in Ref. \cite{Gerber97} in this way. Samples with larger
intergranular couplings did not show any exponential decay but the
resistivity did not saturate at low temperatures and the authors described
its temperature dependence by a power law
\begin{equation}
R=AT^{-\alpha },  \label{a1}
\end{equation}
with $\alpha =0.117$. Apparently, with such a small value of $\alpha $ a
logarithmic temperature dependence
\begin{equation}
R=A\left( 1-\alpha \ln T\right)  \label{b1}
\end{equation}
(obtained by expansion of Eq. (\ref{a1}) in $\alpha )$ could describe the
experimental data as well. It is relevant to emphasize that the array of the
grains was three dimensional and one could not attribute such a behavior to
weak localization effects. \

In this paper, we consider a model for granular metals at not very low
temperature and demonstrate that changing the dimensionless tunneling
conductance $g$ one can have either exponential temperature dependence of
the resistivity at small $g<1$ or the logarithmic behavior, Eq. (\ref{b1}),
at large $g>1$. It will be shown that the result is applicable for any
dimensionality of the array of grains, which contrasts usual logarithmic
corrections due to interference effects \cite{altar} typical for $2D$. The
Hamiltonian describing the model is chosen as
\begin{equation}
\hat{H}=\hat{H}_{0}+\hat{H}_{t}+\hat{H}_{c},  \label{a2}
\end{equation}
where $\hat{H}_{0}$ is the one-electron Hamiltonian of isolated grains
including disorder within the grains. The tunneling of the electrons between
the grains is given by
\begin{equation}
\hat{H}_{t}=\sum_{i,j,\alpha ,\alpha ^{\prime }}t_{ij}\hat{\psi}_{\alpha
i}^{\dagger }\hat{\psi}_{\alpha ^{\prime }j}.  \label{a2a}
\end{equation}
where the summation is performed over the states $\alpha $, $\alpha ^{\prime
}$ of each grain (spin is conserved) and over neighboring grains $i$ and $j$%
. The possibility to tunnel from the state $\alpha $ to an arbitrary state $%
\alpha ^{\prime }$ of other grains introduces an additional disorder
resulting in a finite tunnel conductance.

The term $\hat{H}_{c}$ in Eq. (\ref{a2}) describes the charging energy
\begin{equation}
\hat{H}_{c}=\frac{e^{2}}{2}\sum_{ij}\hat{N}_{i}C_{ij}^{-1}\hat{N}_{j}.
\label{a3}
\end{equation}
In Eq. (\ref{a3}), $\hat{N}_{i}=\sum_{\alpha }\int \hat{\psi}_{\alpha
}^{\dagger }\left( {\bf r}_{i}\right) \hat{\psi}_{\alpha }\left( {\bf r}%
_{i}\right) d{\bf r}_{i}-\bar{N}$ is the excess number of electrons in the $%
i $-th grain. ($\bar{N}$ is the dimensionless local potential) and $C_{ij}$
is the capacitance matrix. Eq. (\ref{a3}) describes the long range part of
the Coulomb interaction in the limit of weak disorder inside the grains and
has been used in many works. Calculations with the Hamiltonian $\hat{H}$,
Eqs. (\ref{a2}, \ref{a3}), can be replaced by computation of a functional
integral over anticommuting fields $\psi _{\alpha }\left( \tau \right) $.

Although the model described by Eqs. (\ref{a2}, \ref{a3}) contains only the
long range part of the Coulomb interaction, it is still very complicated,
because at very low temperatures interference becomes very important and one
has to consider an interplay of localization and interaction effects. One
could do this either using diagrammatic expansions \cite{altar} or writing a
non-linear $\sigma $-model \cite{fin}. Both methods allow to consider the
limit of large tunneling conductances $g$ and the results are strongly
dependent on the dimensionality. However, the behavior, Eq. (\ref{a1}) or
Eq. (\ref{b1}) was not predicted for $3D$ in any of these works.

The model, Eqs. (\ref{a2}, \ref{a3}), becomes simpler if the temperature $T$
is not very low such that low energy diffusion modes are damped. As it was
discussed in a recent publication \cite{belalt}, the granular metal can be
described at temperatures $T\gg g\delta $, where $\delta $ is the mean level
spacing in a single grain, by the Ambegaokar, Eckern and Sch\"{o}n (AES)\cite
{aes} functional of the free energy. If $g\lesssim 1$, this condition should
be replaced by $T\gg \delta $. The limit of not very low temperatures not
only simplifies the consideration but is interesting on its own because it
leads to an unusual behavior of physical quantities and is easily accessible
experimentally. In particular, we will see that changing the tunneling
conductance $g$ one may have a transition from the exponential temperature
dependence of the resistivity to the logarithmic behavior, Eq. (\ref{b1}).

We calculate the conductivity $\sigma \left( \omega \right) $ using the Kubo
formula and making an analytical continuation from Matsubara frequencies $%
i\omega _{n}$ to real frequencies $\omega $ \cite{AGD}. In order to reduce
the calculation of physical quantities to a computation of correlation
functions with the AES action we decouple the interaction term, Eq. (\ref{a3}%
), by integration over an additional $V_{i}\left( \tau \right) $ and then,
following Refs. \cite{gefen,belalt}, remove this field from $\hat{H}_{0}$ by
the gauge transformation
\begin{equation}
\psi _{i}\left( \tau \right) \rightarrow e^{-i\varphi \left( \tau \right)
}\psi _{i}\left( \tau \right) ,\text{ \ \ \ \ \ }\dot{\varphi}_{i}\left(
\tau \right) =V_{i}\left( \tau \right) .  \label{a4}
\end{equation}
This is not a trivial procedure, because the new fields $\psi _{\alpha
}\left( \tau \right) $ should obey, as before, the boundary condition $\psi
_{\alpha }\left( \tau \right) =-\psi _{\alpha }\left( \tau +\beta \right) ,$
$\beta =1/T$. So, the field $V_{i}\left( \tau \right) $ cannot be completely
removed from $\hat{H}_{0}$ at arbitrarily low temperatures. Instead of $%
\varphi _{i}(\tau )$ let us consider $\tilde{\phi}_{i}(\tau )$:
\begin{equation}
\tilde{\phi}_{i}\left( \tau \right) =\phi _{i}\left( \tau \right) +2\pi
Tk_{i}\tau ,  \label{a5}
\end{equation}
where $-\infty <\phi _{i}\left( \tau \right) <\infty $, $\phi _{i}\left(
0\right) =\phi _{i}\left( \beta \right) $, and $k_{i}=0,\pm 1,$ $\pm 2,....$
are the so called winding numbers. Performing the gauge transformation with $%
\tilde{\phi}_{i}$ instead of $\varphi _{i}(\tau )$, the antiperiodicity of
the $\psi _{\alpha }$ is preserved, but the action still contains a term
linear in $i(V_{i}-\dot{\tilde{\phi}}_{i})\in (-i\pi T,i\pi T)$. Only in the
limit $T\gg \delta $ this term can be neglected. The integration over $%
\tilde{\phi}_{i}\left( \tau \right) $ implies integration over $\phi
_{i}\left( \tau \right) $ and summation over $k_{i}$. At large $g\gg 1$, one
can put all $k_{i}=0$. However, at $\ g\lesssim 1$ one should sum over all $%
k_{i}$ and neglecting the contribution of the non-zero winding numbers, as
done in Ref. \cite{gefen}, leads to incorrect results.

Using the phase representation one can write the conductivity $\sigma \left(
\omega \right) $ in the form
\begin{equation}
\sigma \left( \omega \right) =\frac{ia^{d-2}}{\omega }\left[ \int_{0}^{\beta
}d\tau e^{i\Omega _{n}\tau }K\left( \tau \right) \right] _{\Omega
_{n}\rightarrow -i\omega +\delta },  \label{a7}
\end{equation}

\[
K\left( \tau \right) =\langle X_{2}^{{\bf a}}\left( \tau \right) \rangle
-\sum_{{\bf i}}\langle X_{10}^{{\bf a}}\left( \tau \right) X_{1{\bf i}}^{%
{\bf a}}\left( 0\right) \rangle ,
\]
\[
X_{2}^{{\bf a}}\left( \tau \right) =e^{2}\pi g\int_{0}^{\beta }d\tau
^{\prime }\left( \delta \left( \tau \right) -\delta \left( \tau ^{\prime
}-\tau \right) \right) \alpha \left( \tau ^{\prime }\right) \cos \left(
\tilde{\phi}_{{\bf i,i+a}}\left( \tau ^{\prime }\right) -\tilde{\phi}_{{\bf %
i,i+a}}\left( 0\right) \right) ,
\]
\[
X_{1{\bf i}}^{{\bf a}}\left( \tau \right) =e\pi g\alpha \left( \tau -\tau
^{\prime }\right) \sin \left( \tilde{\phi}_{{\bf i,i+a}}\left( \tau ^{\prime
}\right) -\tilde{\phi}_{{\bf i,i+a}}\left( \tau \right) \right) ,\text{ \ \ }%
\alpha \left( \tau \right) =T^{2}\left(
\mathop{\rm Re}%
\left( \sin \left( \pi T\tau +i\delta \right) \right) ^{-1}\right) ^{2}.
\]
where ${\bf a}$ is a vector connecting the centers of neighboring grains $%
{\bf i}$ and ${\bf i}+{\bf a}$, $a=\left| {\bf a}\right| ,$ and $d$ is the
dimensionality of the array. In Eqs. (\ref{a7}), $\tilde{\phi}_{{\bf ij}%
}\left( \tau \right) =\tilde{\phi}_{{\bf i}}\left( \tau \right) -\tilde{\phi}%
_{{\bf j}}\left( \tau \right) $ for ${\bf i}$ and ${\bf j}$ standing for
neighboring grains and
\begin{equation}
\langle ...\rangle =\int \left( ...\right) \exp \left( -S\right) D\tilde{\phi%
}\left( \int \exp \left( -S\right) D\tilde{\phi}\right) ^{-1},  \label{a8}
\end{equation}
where $D\tilde{\phi}$ stands for both the functional integration over $\phi
\left( \tau \right) $ and summation over the winding numbers $k_{i}$. The
AES action $S$ can be written as
\begin{equation}
S=S_{c}+S_{t},  \label{a9}
\end{equation}
where $S_{c}$ describes the charging energy
\begin{equation}
S_{c}=\frac{1}{2e^{2}}\sum_{{\bf ij}}\int_{0}^{\beta }d\tau C_{ij}\frac{d%
\tilde{\phi}_{i}\left( \tau \right) }{d\tau }\frac{d\tilde{\phi}_{j}\left(
\tau \right) }{d\tau }  \label{a10}
\end{equation}
and $S_{t}$ stands for tunneling between the grains
\begin{equation}
S_{t}=2\pi g\sum_{\left| {\bf i}-{\bf j}\right| =a}\int_{0}^{\beta }d\tau
d\tau ^{\prime }\alpha \left( \tau -\tau ^{\prime }\right) \sin ^{2}\left(
\frac{\tilde{\phi}_{ij}\left( \tau \right) -\tilde{\phi}_{ij}\left( \tau
^{\prime }\right) }{2}\right)  \label{a11}
\end{equation}
The dimensionless conductance $g$ is given by $g=2\pi \nu ^{2}t_{ij}^{2}$,
where $t_{ij}$ is the tunneling amplitude from grain $i$ to grain $j$ (spin
is included).

Although the model described by Eqs. (\ref{a7}-\ref{a11}) is simpler than
the initial model, Eqs. (\ref{a2}-\ref{a3}), explicit formulae can be
written only in limiting cases. The same action $S$, Eqs. (\ref{a9}-\ref{a11}%
), was used in Ref. \cite{fazio}, and a metal-insulator transition has been
predicted in a $2D$ array of tunnel junctions. However, the authors of Ref.
\cite{fazio} did not calculate the conductivity but discussed properties of
the partition function. For large $g$ they did not account for phase
fluctuations properly which, as we show here, are responsible for the
behavior, Eq. (\ref{b1}). Moreover, we find a transition in any
dimensionality.

If the temperature $T$ is very high, $T\gg E_{c}\sim e^{2}C_{ij}^{-1}$,
where $E_{c}$ is the electrostatic energy of adding one electron to a grain,
fluctuations of the phases $\tilde{\phi}$ are negligible and one can set $%
\tilde{\phi}=0$ in the expressions for $X_{1\text{ }}$and $X_{2}$ in Eqs. (%
\ref{a7}). Then, we obtain easily the conductivity
\begin{equation}
\sigma _{0}=e^{2}ga^{d-2},  \label{a12}
\end{equation}
which shows that at such temperatures charging interactions are not
important.

In the opposite limit $T\ll E_{c},$ transport in the granulated system has
much more interesting characteristics. This inequality can be compatible
with the inequality $T\gg \max \{g\delta ,\delta \},$ used for the
derivation of Eqs. (\ref{a7}-\ref{a11}), because $E_{c}\gg \delta $ for $2D$
and $3D$ grains.

We calculate the conductivity at temperatures $T\ll E_{c}$ in the limits $%
g\gg 1$ and $g\ll 1$.

In the limit of large conductances $g\gg 1$, fluctuations of $\phi $ are
small and all non-zero winding numbers $k_{i}$ can be neglected. Non-zero $%
k_{i}$ (as well as variations of $\bar{N}_{i}$ ) would lead to contributions
of order $\exp \left( -g\right) ,$ and can be neglected in any expansion in $%
1/g$.

Keeping only quadratic in $\phi $ terms in Eqs. (\ref{a9}-\ref{a11}) we
reduce the action $S$ to the form
\begin{equation}
S=T\sum_{{\bf q},n}\phi _{{\bf q},n}G_{{\bf q},n}^{-1}\phi _{-{\bf q},-n},
\label{a13}
\end{equation}
\[
G_{{\bf q},n}^{-1}=\omega _{n}^{2}/\left( 4E\left( {\bf q}\right) \right)
+2g\left| \omega _{n}\right| \sum_{{\bf a}}\left( 1-\cos {\bf q\bar{a}}%
\right) ,
\]
where $E\left( {\bf q}\right) =e^{2}/\left( 2C\left( {\bf q}\right) \right) $
and $C\left( {\bf q}\right) $ is the Fourier-transform of the capacitance
matrix $C_{ij}$ (${\bf q}$ are quasi-momenta for the array of the grains).
One should sum in Eq. (\ref{a13}) over $d$ unit lattice vectors ${\bf \bar{a}%
,}$ where $d$ is the dimensionality of the array.

Keeping only quadratic in $\phi $ terms in the action but not expanding the
function $X_{2}$, Eq. (\ref{a7}), one reduces the correlator $\langle X_{2%
{\bf a}}\left( \Omega _{n}\right) \rangle $ to the form
\[
\langle X_{2{\bf a}}\left( \Omega _{n}\right) \rangle =\pi
e^{2}g\int_{0}^{\beta }\alpha \left( \tau \right) \left( 1-e^{i\Omega
_{n}\tau }\right) e^{-\tilde{G_{{\bf a}}}\left( \tau \right) }d\tau ,
\]

\begin{equation}
\tilde{G}_{{\bf a}}\left( \tau \right) =4T\int \frac{d{\bf q}}{\left( 2\pi
\right) ^{d}}G_{{\bf q}n}\sin ^{2}\frac{{\bf q\bar{a}}}{2}\sin ^{2}\frac{%
\omega _{n}\tau }{2}  \label{a14}
\end{equation}

One can check that the contribution coming from the correlator of the
functions $X_{1}$ in Eqs. (\ref{a7}) contains additional powers of $1/g$ and
can be neglected in the main approximation. It is very important that the
correlator $\langle X_{1}X_{1}\rangle $ in Eqs. (\ref{a7}) contains a
summation over $j,$ which corresponds to the zero quasi-momentum of the
function $K$. If we carried out the computation for a single grain the
contribution from $\langle X_{1}X_{1}\rangle $ would not be smaller than the
one from $\langle X_{2}\rangle $.

What remains to be done in order to calculate the conductivity for $g\gg 1$
is to compute the integral in Eq. (\ref{a14}) for the Matsubara frequencies $%
\Omega _{n}$ and make the analytical continuation $\Omega _{n}\rightarrow
-i\omega +\delta $. In the lowest order in $\alpha $ the result for the
conductivity $\sigma $ in the limit $\omega \rightarrow 0$ is
\begin{equation}
\sigma =\sigma _{0}\left( 1-\alpha \ln \left( gE_{c}/T\right) \right) ,\text{
\ \ \ \ \ }\alpha =\left( 2\pi gd\right) ^{-1}.  \label{a15}
\end{equation}

Thus, at large $g\gg 1$ the conductivity decays with temperature
logarithmically. Of course, one may not use this formula at very low
temperature because the present consideration is valid at not very low
temperatures $T\gg g\delta $ when the AES action may be used. At lower
temperatures, one should take into account interference effects and,
depending on the dimensionality $d$ of the array, both metal and insulating
states are possible. In contrast, Eq. (\ref{a15}), is valid in {\em any }%
dimensionality.

If we used Eq.(\ref{a14}) exactly we would obtain the power law, Eq. (\ref
{a1}), with the exponent $\alpha $ from Eq. (\ref{a15}). A similar
dependence was written for the voltage dependence of the conductance of a
single junction in a model with an electromagnetic environment \cite
{devoret,girvin}. However, taking into account non-quadratic terms in the
expansion of the action in phases $\phi $ changes this result and one comes
again to Eq. (\ref{a15}). For a single junction this result is known since
the works of Ref. \cite{schmid} where a proper renormalization group (RG)
equation was written. One can check that, in the first order of the RG, the
equation is the same for a granular metal and does not depend on the
arrangement of the grains in the array. It can be written as
\begin{equation}
\frac{dg\left( \xi \right) }{d\xi }=-\frac{1}{2\pi d}  \label{b15}
\end{equation}
where $\xi =-\ln \tau $ and $g\left( \xi \right) $ is the effective
conductance.

Solving Eq. (\ref{b15}) with the boundary condition $g\left( 0\right) =g$ we
come immediately to Eq. (\ref{a15}).

However, next orders in the RG (expansion of the Gell-Mann-Low function in $%
g\left( \xi \right) $) are dependent on the type of the array and differ
from the corresponding terms for a single junction. The applicability of the
one-loop approximation, Eq. (\ref{b15}), and of its solution, Eq. (\ref{a15}%
), implies an additional inequality for the temperature: $T\geq
T_{c}=gE_{c}\exp (-1/\alpha )$. In other words, Eq. (\ref{a15}) remains
valid until $\sigma /e^{2}$ becomes of the order unity. For not very large
grains this inequality is less restrictive because $T_{c}$ is exponentially
small for large $g.$ The dimensionality of the array $d$ enters Eqs. (\ref
{a15}, \ref{b15}) as a parameter only. Actually, the number of contacts with
neighboring grains (coordination number) rather than the dimensionality
itself enters Eqs. (\ref{a15}, \ref{b15}). This difference may be important
in situations when the grains are close packed in a cubic lattice.

The logarithmic behavior, Eq. (\ref{a15}), describes the granular system at
sufficiently large $g\gtrsim 1$. At smaller $g$, the temperature dependence
becomes exponential and we check this statement in the limit $g\ll 1$
expanding the functional integral in Eq. (\ref{a8}) in the tunneling part $%
S_{t}$, Eq. (\ref{a11}), of the action. The main contribution comes again
from the function $\langle X_{2}\left( \tau \right) \rangle $ in Eqs. (\ref
{a7}). In the lowest order one can completely neglect $S_{t},$ which leads
to computation of the correlator $\Pi \left( \tau \right) $
\begin{equation}
\Pi \left( \tau \right) =\left\langle \exp \left( -i\left( \tilde{\phi}%
_{i}\left( \tau \right) -\tilde{\phi}_{i}\left( 0\right) \right) \right)
\right\rangle _{S_{c}},  \label{a16}
\end{equation}
where the phases $\tilde{\phi}_{i}\left( \tau \right) $ are introduced in
Eq. (\ref{a5}), $S_{c}$ is given by Eq. (\ref{a10}) and the averaging should
be performed with this functional (Strictly speaking the function $\Pi
\left( \tau \right) $ is sufficient for calculating $K\left( \tau \right) $
only for diagonal $C_{ij}$. However, a proper modification for an arbitrary $%
C_{ij}$ is simple). The computation of the average in Eq. (\ref{a16}) can be
performed using two different methods. A more straightforward way of
calculating is to use the definition of $\tilde{\phi}_{i}\left( \tau \right)
$, Eq. (\ref{a5}), which allows to represent the action $S_{c}$ as $%
S_{c}=S_{c}[\phi ]+S_{c}[k]$ and carrying out integration over the phase $%
\phi $ and summation over the winding numbers separately. Integrating over
the phase $\phi _{i}\left( \tau \right) $ we obtain for $0<\tau <\beta $
(see also \cite{fazio,gefen})
\begin{equation}
\langle e^{-i\left( \phi _{i}\left( \tau \right) -\phi _{i}\left( 0\right)
\right) }\rangle =\exp \left( -B_{ii}\left( \tau -T\tau ^{2}\right) \right) ,
\label{a19}
\end{equation}
where $B_{ij}=\frac{e^{2}}{2}\left( C^{-1}\right) _{ij}$.

However, Eq. (\ref{a19}) is not the final result and the summation of the
winding numbers is essential. This can be performed using the Poisson
summation formula. As a result, the function $\Pi \left( \tau \right) $ can
be represented as
\begin{equation}
\Pi \left( \tau \right) =\frac{1}{Z}e^{-B_{ii}\tau
}\sum_{\{n_{k}\}}e^{-\sum_{k}2\tau n_{k}B_{ki}-\beta
\sum_{k,l}B_{kl}n_{k}n_{l}},  \label{a20}
\end{equation}
where all $n_{k}$ are integers and $Z$ is a normalization coefficient ($\Pi
\left( 0\right) =1$). The necessary periodicity in $\tau $ of the function $%
\Pi \left( \tau \right) $ with the period $\beta $ is evident from Eq.(\ref
{a20}).

The second method is to use the standard quantum mechanical formalism
instead of calculating the functional integrals in Eq. (\ref{a16}), which
has been suggested in an earlier work on granulated superconductors \cite
{efetov80}. Within this approach one writes instead of the action $S_{c}$,
Eq. (\ref{a10}), an effective Hamiltonian $\hat{H}_{{\rm eff}}$%
\begin{equation}
\hat{H}_{{\rm eff}}=\sum_{ij}B_{ij}\hat{\rho}_{i}\hat{\rho}_{j}\text{, \ \ \
}\hat{\rho}_{i}=-i\partial /\partial \phi _{i},  \label{a21}
\end{equation}
and calculates the thermodynamic average with $\hat{H}_{{\rm eff}}$. For the
function $\Pi \left( \tau \right) $ one should calculate the average
\[
\langle e^{-i\left( \hat{\phi}_{i}\left( \tau \right) -\hat{\phi}_{i}\left(
0\right) \right) }\rangle _{\hat{H}_{{\rm eff}}}\text{, \ \ }\hat{\phi}%
_{i}\left( \tau \right) =e^{\hat{H}_{{\rm eff}}\tau }\phi _{i}e^{-\hat{H}_{%
{\rm eff}}\tau }.
\]
Eigenvalues of the operators $\hat{\rho}_{i}$ are integers (eigenfunctions
of $H_{{\rm eff}}$ must be periodic in $\phi $ with the period $2\pi $) and
one comes easily to Eq. (\ref{a20}). This consideration explicitly
demonstrates that accounting for the winding numbers leads to the charge
quantization.

\ We see from Eq. (\ref{a20}) that, in order to get an explicit expression
for the conductivity, one should sum over all configurations of charge. At
high temperatures $T\gg E_{c}$ the sum over $n_{k}$ in Eq. (\ref{a20}) can
be replaced by integrals and we get $\Pi \left( \tau \right) =1$, which
leads to Eq. (\ref{a12}).

In the opposite limit, $T\ll E_{c}$, the main contribution comes from charge
configurations with the lowest energy. The ground state with all $n_{k}=0$
does not contribute to the conductivity. If the lowest excited state
corresponds to one charged grain with the charge $\pm 1$ (depending on a
particular $C_{ij}$) we come using Eqs. (\ref{a7}) in the limit $\omega
\rightarrow 0$ to a rather simple formula
\begin{equation}
\sigma =2\sigma _{0}\exp \left( -B_{ii}/T\right) .  \label{a22}
\end{equation}

Since $B_{ii}$ is the energy corresponding to the charge $\pm 1$, Eq. (\ref
{a22}) corresponds to conduction of an activated electron and hole (the
factor $2$ means that both of them are taken into account).

Comparing Eq. (\ref{a22}) with Eq. (\ref{a15}) we come to the conclusion
that there must be a critical value $g_{c}$ separating in the limit $%
T\rightarrow 0$ the logarithmic behavior from the exponential one. Whether
the activation energy (Coulomb gap) turns to zero or has a jump at $g=g_{c}$
is not clear from the present consideration. (Strictly speaking, there
should not be any singularity of the conductivity at finite temperatures but
the change of the behavior may be noticeable experimentally or numerically).
The model of the granular metal may be used to describe disordered electron
systems at low electron density. In this case, potential wells would
correspond to the grains.

The sample of the experiment \cite{Gerber97} that showed the ``power law
behavior'', Eq. (\ref{a1}), had the room temperature resistivity $%
R_{0}=7.3\times 10^{-3}\Omega $cm. The diameter of the grains was $120\pm 20$%
\AA $,$ which allows, using the value $\hbar /e^{2}=4.1\times 10^{3}\Omega $%
, to estimate the dimensionless tunnel conductivity as $g=0.7$. If we put $%
d=2$ in Eq. (\ref{a15}) we obtain $\alpha =0.116$, which exactly corresponds
to the experimental value from Eq. (\ref{a1}). However, the arrays used in
Ref. \cite{Gerber97} were rather thick and, at first glance, one should use $%
d=3$. Nevertheless, the value of $d$ in Eq. (\ref{a15}) corresponds rather
to the half of the contacts of a single grain than to the real
dimensionality. Then, the experimental value of $\alpha $ indicates that
either the grains are not closely packed such that the typical number of
contacts per grain is $4$ or our calculation is too rough to provide a
quantitative agreement with the experiment (the value of $\alpha $, Eq. (\ref
{a15}), is based on the assumption $g\gg 1$ but the experimental value of $g$
is of order $1$). The resistivity of samples with a high room temperature
resistivity behaved as $\exp \left( a/T^{1/2}\right) $ rather than obeying
the activation law, Eq. (\ref{a22}). But this can be attributed to a
variation of the size of the grains or of the local potential \cite{abeles}.

In conclusion, we suggested a scheme of calculating the conductivity of a
granular metal at not very low temperatures. On the basis of explicit
results we demonstrated the existence in any dimensionality of a transition
between states with exponential dependence of the conductivity on
temperature and a logarithmic one. Relating the coefficient $\alpha $ to the
room temperature conductivity we were able to compare our results with an
existing experiment and got a good agreement. The model of the granular
metal may serve also as a good description of disordered systems with a low
electron density. \acknowledgments
We are grateful to A. Altland and B.L. Altshuler, A.V. Andreev, L.I.
Glazman, A.I. Larkin, and Yu.V. Nazarov for discussions. A support of the
{\it Sonderforschungsbereich} 237 is greatly appreciated.

\end{document}